\documentclass[aps,prl,showpacs,twocolumn,superscriptaddress,
              floats,floatfix]{revtex4}
\usepackage{amssymb,amsmath}
\usepackage{graphics}
\usepackage{color}
\usepackage{subfigure}
\usepackage{epsfig}
\begin{document}
\title{Statistically Steady Turbulence in Soap Films:
Direct Numerical Simulations with Ekman Friction}
\author{Prasad Perlekar}
\email{perlekar@physics.iisc.ernet.in}
\affiliation{Centre for Condensed Matter Theory, Department of Physics, Indian
Institute of Science, Bangalore 560012, India.} 
\author{Rahul Pandit}
\email{rahul@physics.iisc.ernet.in}
\altaffiliation[\\ also at~]{Jawaharlal Nehru Centre For Advanced
Scientific Research, Jakkur, Bangalore, India}
\affiliation{Centre for Condensed Matter Theory, 
Department of Physics, 
Indian Institute of Science, Bangalore 560012, India.} 
\begin{abstract}

We present a detailed direct numerical simulation (DNS) designed to 
investigate the combined effects of walls and Ekman friction  on 
turbulence in forced soap films. We concentrate on the forward-cascade 
regime and show how to extract the isotropic parts of velocity 
and vorticity structure functions and thence the ratios of 
multiscaling exponents. We find that velocity structure functions 
display simple scaling whereas their vorticity counterparts show 
multiscaling; and the probability distribution function of the Weiss parameter 
$\Lambda$, which distinguishes between regions with centers and 
saddles, is in quantitative agreement with experiments.

\end{abstract}
\maketitle

The pioneering work of Kraichnan~\cite{kraic67} showed that fluid turbulence 
in two dimensions (2D) is qualitatively different from that in three 
dimensions (3D): in the former we have an infinity of extra conserved 
quantities, in the inviscid, unforced case; the first of these is the 
enstrophy.  It turns out, therefore, that 2D turbulence displays an inverse 
cascade of energy, from the length scale at which the force acts to larger 
length scales, and a forward cascade of enstrophy, from the forcing length 
scale to smaller ones; by contrast, 3D turbulence is characterised by a 
forward cascade of energy~\cite{Fri96}.  Kraichnan's predictions were first 
confirmed in atmospheric experiments in quasi-two-dimensional, stratified 
flows~\cite{boe83};  subsequent experiments have studied systems ranging 
from large-scale geophysical flows to soap 
films~\cite{boe83,par02,riv00,riv01,dan02,riv07}. The 
latter have proved to be especially useful in characterizing 2D turbulence. 

We present the first direct numerical study (DNS) that has been 
designed specifically to explore the combined effects of the 
air-drag induced Ekman friction $\alpha$ and walls on the forward cascade in 2D 
turbulence; and we employ the Kolmogorov 
forcing used in many soap-film experiments~\cite{riv00,riv01,dan02,riv07}.  
Thus we can make a far more detailed comparison with these experiments 
than has been attempted hitherto. In particular, if we use values 
of $\alpha$ that are comparable to those in experiments, we find 
that the energy dissipation rate because of the Ekman friction is 
comparable to the energy dissipation rate that arises from  the 
conventional viscosity. We show how to extract the 
isotropic parts~\cite{bou05} of velocity and vorticity structure 
functions and then, by using the extended self-similarity (ESS) 
procedure~\cite{ben93}, we obtain ratios of multiscaling 
exponents whence we conclude that velocity structure functions 
show simple scaling whereas their vorticity counterparts display  
multiscaling. Most important, our probability distribution function~(PDF) 
of the Weiss parameter $\Lambda$~\cite{wei92} is in quantitative agreement 
with that found in experiments~\cite{riv01,dan02}.

For the low-Mach-number flows we consider, we can use the following soap-film 
equations~\cite{cho01,riv00}: 
\begin{eqnarray} 
(\partial_t + {\bf u}\cdot \nabla) {\omega} = \nu \nabla^2 {\omega} - \alpha {\omega} + 
F_{\omega}/\rho;\;\;\; \nabla^2 \psi = \omega.
\label{eq:ns_vor} 
\end{eqnarray} 
Here ${\bf u}\equiv(-\partial_y \psi, \partial_x \psi)$, $\psi$, and 
$\omega\equiv \nabla \times {\bf u}$ are, respectively, the velocity, 
stream function, and vorticity at the position ${\bf x}$ and 
time $t$; we choose the uniform density $\rho=1$; $\alpha$ is the 
Ekman friction coefficient, $\nu$ is the kinematic viscosity, 
and $F_{\omega} \equiv k_{inj} F_0 \cos(k_{inj}x)$, a Kolmogorov-type 
forcing term, with amplitude $F_0$, and injection wave vector $k_{inj}$ 
(the length scale $\ell_{inj}\equiv 2\pi/k_{inj})$. We impose 
no-slip ($\psi=0$) and no-penetration ($\nabla\psi \cdot \hat{n} =0$) 
boundary conditions on the walls, where $\hat{n}$ is the outward normal 
to the wall. If we non-dimensionalize ${\bf x}$ by $k_{inj}^{-1}$, 
$t$ by $k_{inj}^{-2}/\nu$, and $F_\omega$ by $2\pi/(k_{inj} || F_\omega||_2)$, 
with $||F_\omega||_2 \equiv (\int_{A} |F_\omega|^2d{\bf x})^{1/2}$, 
then we have two control parameters, namely, the 
Grashof~\cite{Doe95} number ${\mathcal G}=2\pi||F_\omega||_2/(k_{inj}^{3}\rho\nu^2)$ 
and the non-dimensionalized Ekman friction $\gamma=\alpha/ (k_{inj}^2\nu)$. 
For a given set of values of ${\mathcal G}$ and $\gamma$, the system attains 
a nonequilibrium statistical steady state after a time $t/\tau \simeq 2.8$, 
where $\tau = L/u_{rms}$ is the box-size time, $L$ the side of our square 
simulation domain, and $u_{rms}$ the root-mean-square velocity. In this 
state the Reynolds number $Re\equiv u_{rms}/(k_{inj}\nu)$, the energy, etc., 
fluctuate; their mean values, along with one-standard-deviation error bars, 
are given in Table~\ref{table:para} that lists the values of the parameters 
in our runs {\tt R1-7}.

We use a fourth-order Runge-Kutta scheme with step size $\delta t = 10^{-4}$ 
for time marching in Eq.~\eqref{eq:ns_vor} and evaluate spatial 
derivatives via second-order and fourth-order, centered, finite differences, 
respectively, for points adjacent to the walls and for points inside the 
domain. The Poisson equation in~\eqref{eq:ns_vor} is solved by using a 
fast-Poisson solver~\cite{nr} and $\omega$ is calculated at the boundaries 
by using Thom's formula~\cite{wei96}.  To evaluate spatiotemporal 
averages, we store $\psi({\bf x},t_n)$ and $\omega({\bf x},t_n)$, with 
$t_n= (4 + n \Delta)\tau$, $n = 0, 1, 2, \ldots, n_{max}$, and 
$96 \leq n_{max}\leq 200$; $\Delta=0.28$ for runs {\tt R1-6} and 
$\Delta=0.13$ for run {\tt R7}. 

\begingroup \squeezetable \begin{table*}
%%%%%%%%%%%%%%%% TABLE OF PARAMETERS  %%%%%%%%%%%%%%%%%
\begin{tabular}{@{\extracolsep{\fill}} c c c c c c c c c c c c c
c} \hline $ $ &$N$   & $\alpha$ & $F_0$ & $\gamma$ & ${\mathcal
G}(\times 10^4)$ & $Re$ & $E$ & $\epsilon_{\nu}$ & $\epsilon_e$ &
$\tilde{\Lambda}(\times 10^2)$ & $b$ & $\delta_b(\times 10^{-2})$
\\ \hline \hline ${\tt R1}$  & $1025$  & $0.45$ & $45$ &$0.25$ &
$3.5$ & $23.3\pm0.4$ & $15.1\pm0.5$ & $-28\pm2$ & $-13.6\pm0.5$ &
$5.3\pm0.3$ & $0.32\pm0.01$ & $3.1\pm 0.1$\\ \hline	${\tt
R2}$  & $1025$  & $1.25$ & $45$ & $0.71$ & $3.5$ & $19.6\pm0.3$ &
$10.7\pm0.3$ & $-28\pm1$ & $-26.8\pm0.9$ & $4.8\pm0.2$ &
$0.33\pm0.01$ & $3.1\pm 0.1$\\ \hline	${\tt R3}$  & $1025$  &
$1.25$ & $60$ & $0.71$ & $4.7$ & $24.0\pm0.5$ & $15.9\pm0.6$ &
$-40\pm2$ & $-39.9\pm1.4$ & $7.2\pm0.4$ & $0.33\pm0.01$ &
$2.8\pm0.1$ \\ \hline	${\tt R4}$  & $2049$  & $0.45$ & $45$ &
$0.25$ & $3.5$ & $23.2\pm0.4$ & $15.1\pm0.5$ & $-28\pm2$ &
$-13.6\pm0.4$ & $5.3\pm0.3$ & $0.31\pm0.01$ & $3.2\pm0.2$\\
\hline	${\tt R5}$  & $2049$  & $1.25$ & $45$ & $0.71$ & $3.5$ &
$19.6\pm0.4$ & $10.8\pm0.4$ & $-28\pm1$ & $-27.0\pm1.0$ &
$4.8\pm0.2$ & $0.33\pm0.01$ & $3.1\pm0.1$\\ \hline	${\tt
R6}$  & $2049$  & $1.25$ & $60$ & $0.71$ & $4.7$ & $23.8\pm0.4$ &
$15.9\pm0.6$ & $-40\pm2$ & $-40.0\pm1.5$ & $7.2\pm0.4$ &
$0.33\pm0.01$ & $3.1\pm0.1$\\ \hline ${\tt R7}$  & $3073$  &
$0.45$ & $45$ & $0.25$ & $3.0$ & $26.5\pm0.4$ & $20.0\pm1.0$ &
$-26\pm2$ & $-17.8\pm0.6$ & $5.0\pm0.4$ & $0.31\pm0.01$ & $3.7\pm
0.3$ \\ 
\hline 
\end{tabular} 
\caption{ Parameters for our runs {\tt R1-7}: $N$, the number of grid points 
along each direction, $\gamma$, ${\mathcal G}$, $Re$ (we use $\nu=0.016$,
$\ell_{inj}=0.6$, and a square simulation domain with side $L=7$, 
grid spacing $\delta_x = L/N$, area $A$, and boundary $\partial A$), 
the time-averaged kinetic energy, viscous-energy-dissipation
rate, and the energy-dissipation rate because of Ekman friction, 
$E,\,\epsilon_{\nu}$, and $\epsilon_e$, respectively, 
$\tilde{\Lambda}\equiv\sqrt{\langle(\partial_x u_y^\prime)^2\rangle\langle(\partial_y u_x^\prime)^2\rangle}$, 
$b\equiv-\langle\partial_xu_y^\prime \partial_y u_x^\prime \rangle/\tilde{\Lambda}$, and  the 
boundary-layer thickness $\delta_b \propto\langle(\oint_{\partial A} \omega^2 / \oint_{\partial A} 
((\nabla\omega).\hat{n})^2)^{1/2}\rangle$~\cite{cle05}.}  
\label{table:para}
\end{table*} 
\endgroup

\begin{figure}
\includegraphics[width=0.86\linewidth]{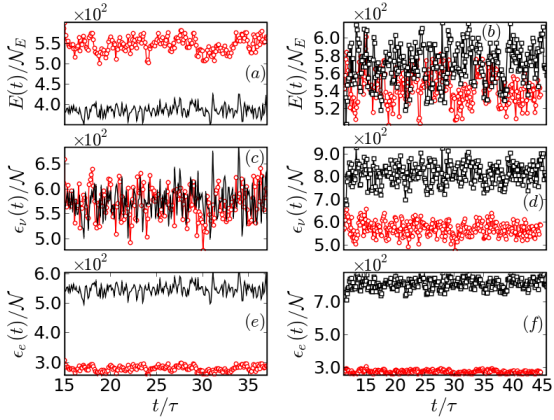}
\caption{\small 
(Color online)  Representatitve plots from runs  ${\tt R1}$(red circles), 
${\tt R2}$(black lines),  and ${\tt R3}$(black squares), showing the time 
evolution of  $E(t)/{\mathcal N}_E$~[$(a)$ and $(b)$], 
$\epsilon_\nu(t)/{\mathcal N}$~[$(c)$ and $(d)$], and
$\epsilon_e(t)/{\mathcal N}$~[$(e)$ and $(f)$]. In $(a)$,
$(c)$, and $(e)$ we keep ${\mathcal G}$ fixed and  vary
$\gamma$~($\gamma=0.25$(red circles) and
$\gamma=0.71$(black line)). In $(b)$, $(d)$, and $(f)$ we
maintain $Re\simeq21.2$ by varying $\gamma$~($\gamma=0.25$(red 
circles) and $\gamma=0.71$(black squares)) and ${\mathcal G}$.
}
\label{fig:eevol}
\end{figure}

Figures~\ref{fig:eevol}(a)-(f) show the time evolution of the kinetic energy 
$E(t)\equiv(\int_{\bf A} {\bf u}^2 d{\bf x})/A$, viscous energy-dissipation 
rate $\epsilon_{\nu}(t) \equiv -\nu (\int_{\bf A} |\omega|^2 d{\bf x})/A$, and 
energy-dissipation rate because of the Ekman friction 
$\epsilon_{e}(t)=-2\alpha E(t)$ (non-dimensionalized, respectively, by 
${\mathcal N}_E \equiv (\nu k_{inj})^2$ and ${\mathcal N} \equiv -k_{inj}^4\nu^3$).  
The mean values $E\equiv \langle E(t) \rangle$, $\epsilon_{\nu} \equiv \langle\epsilon_{\nu}(t) \rangle$, 
and $\epsilon_{e} \equiv \langle \epsilon_{e}(t)\rangle$, given in 
Table \ref{table:para}, are comparable to those in experiments; note that 
$\epsilon_\nu$ and $\epsilon_e$ are of similar magnitudes. By comparing data 
from runs ${\tt R1}$ (red circles) and ${\tt R2}$ (black lines) in 
Figs.~\ref{fig:eevol}(a), (c), and (e) we see that, if we fix ${\mathcal G}$ and 
increase $\gamma$, $E$ decreases, $\epsilon_\nu$ remains unchanged (within error bars), 
and $\epsilon_e$ increases. If we change both ${\mathcal G}$ and $\gamma$, we can keep the 
mean $Re$ fixed, as in runs {\tt R1} and {\tt R3} in Table~\ref{table:para}, by compensating 
an increase in $\gamma$ with an increase in ${\mathcal G}$ (cf. Ref.~\cite{riv01}); 
in Figs.~\ref{fig:eevol}(b), (d), and (f) we see, by comparing runs {\tt R1} (red circles) and {\tt R3}
(black squares), that $E$ remains unchanged (within error bars), whereas both 
$\epsilon_\nu$ and $\epsilon_e$ increase as $\gamma$ and ${\mathcal G}$ increase in such a way 
that $Re$ is held fixed.

\begin{figure*}
\includegraphics[width=0.3\linewidth]{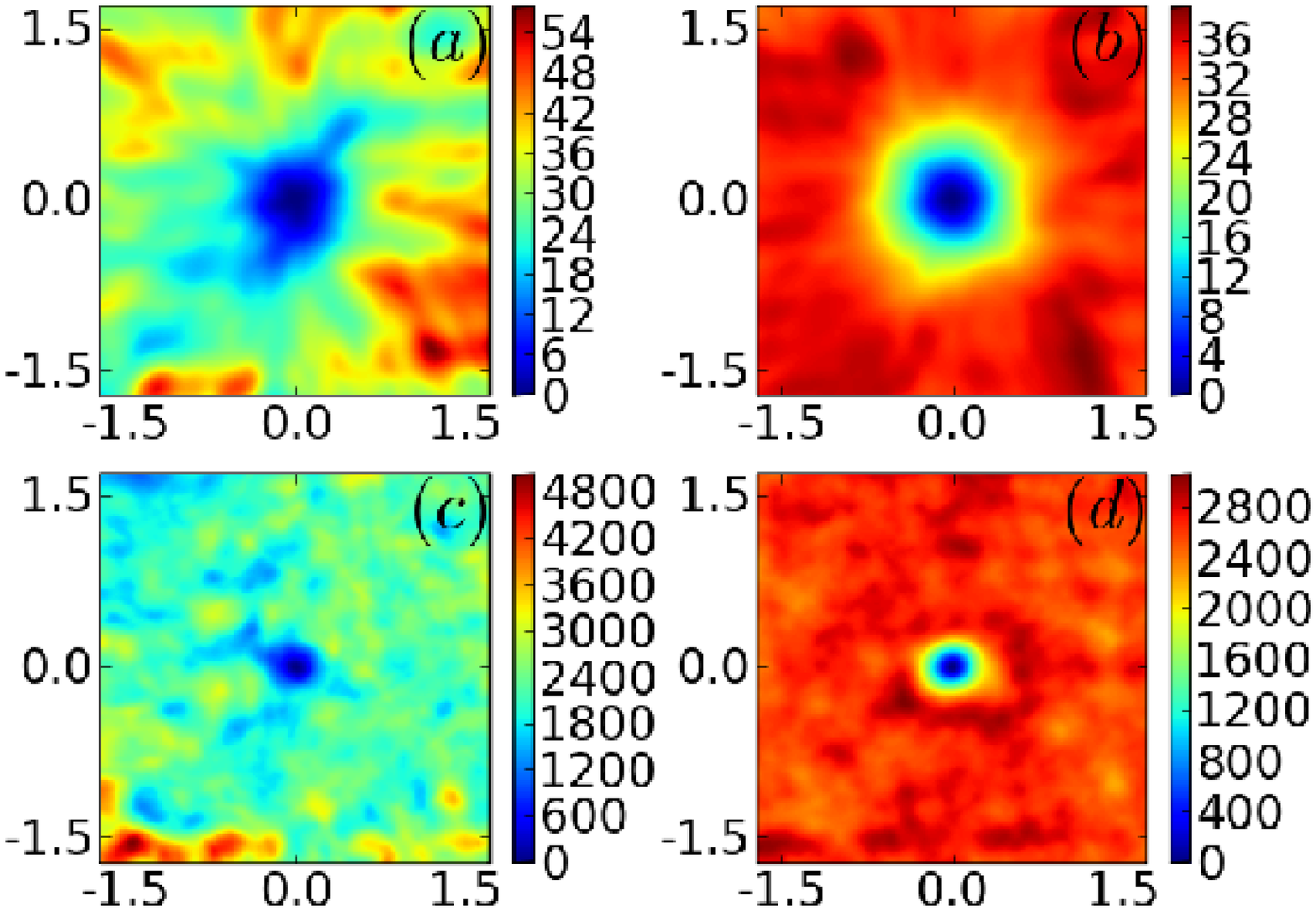}
\includegraphics[width=0.3\linewidth]{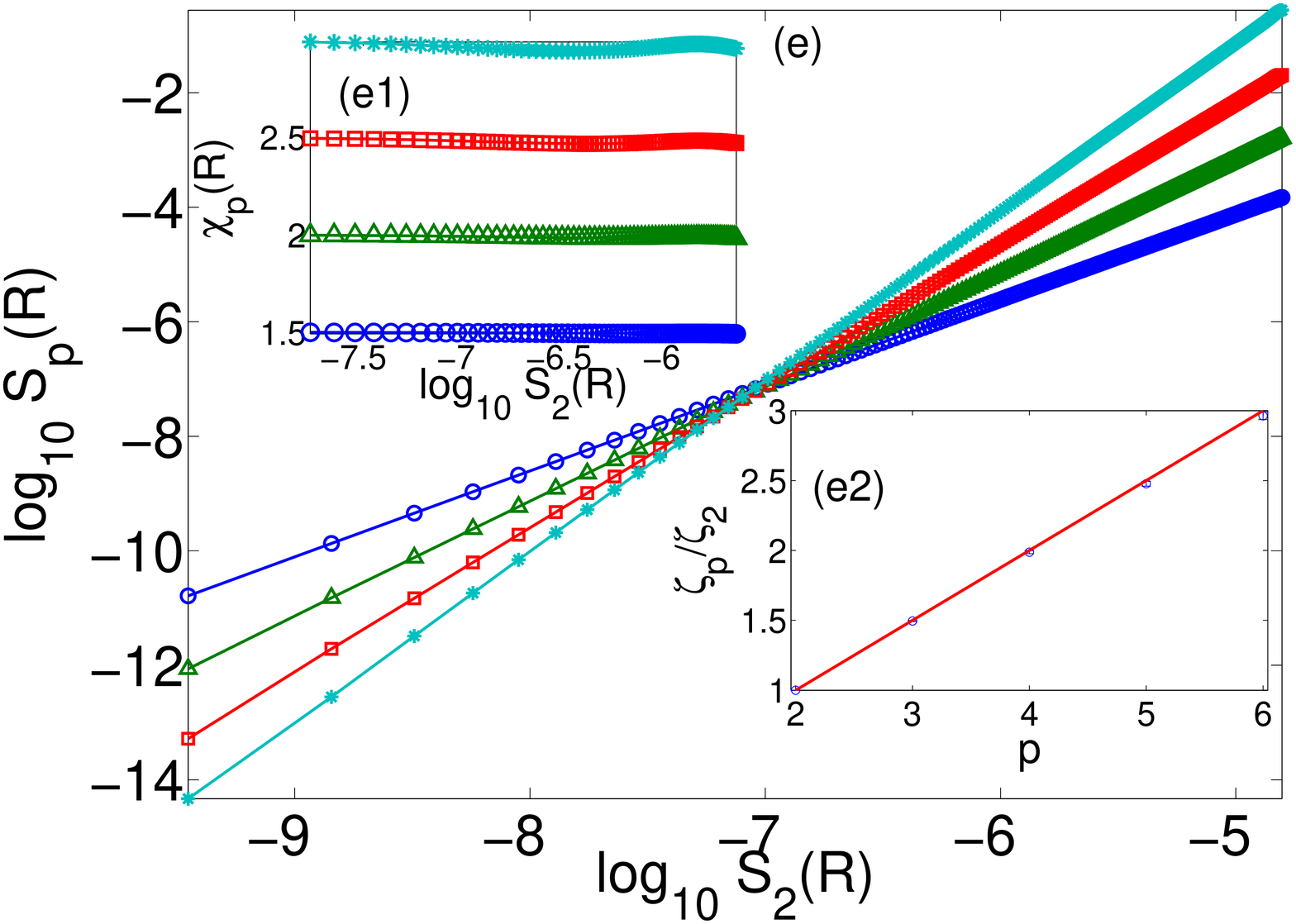}
\includegraphics[width=0.3\linewidth]{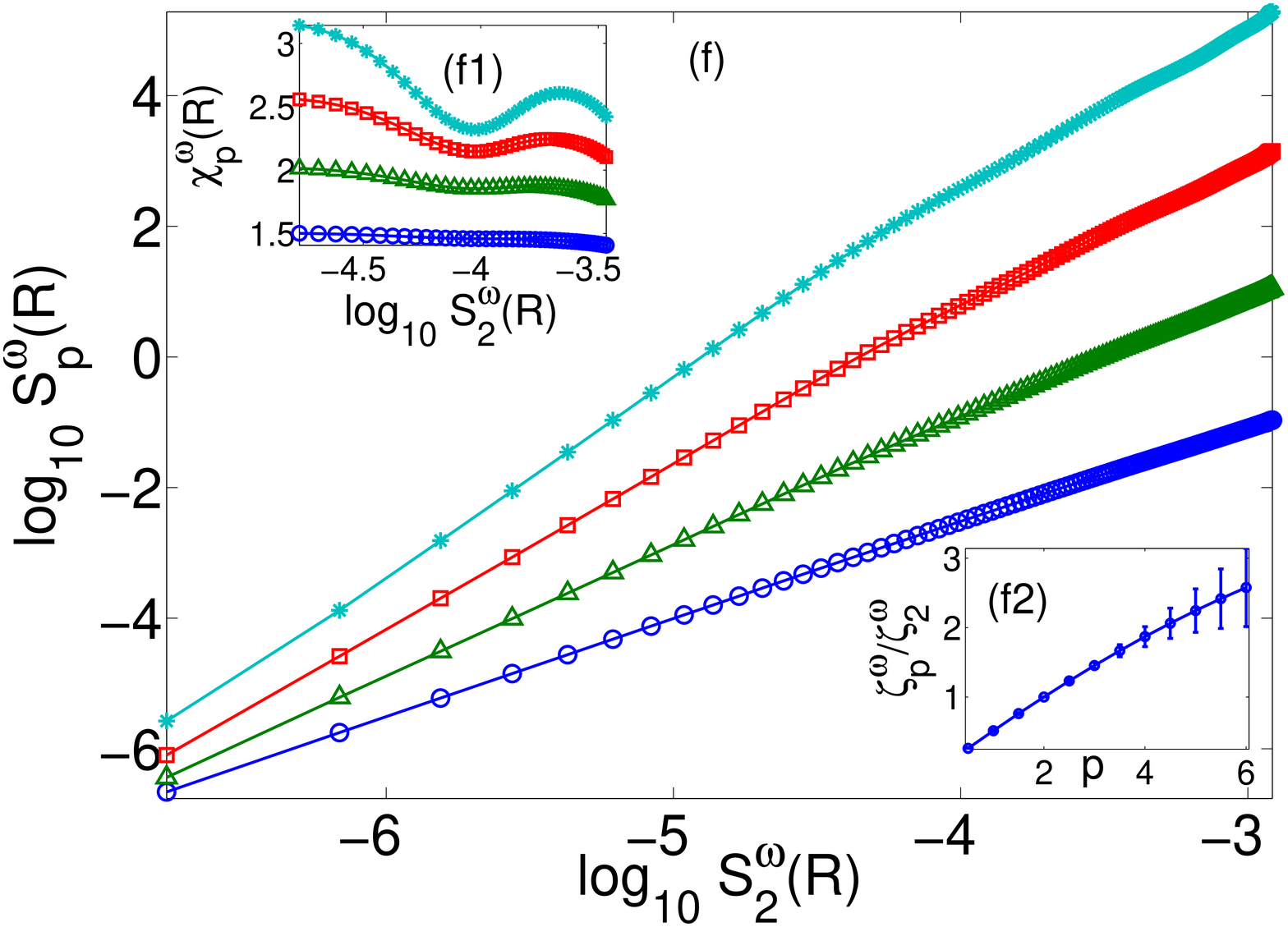}
\caption{\small 
(Color online) Pseudocolor plots of (a) $S_2({\bf r_c,R})$, for 
${\bf r_c}=(2,2)$, (b) $S_2(R)$ (average of $S_2({\bf r_c,R})$ 
over ${\bf r_c}$), (c) $S^{\omega}_2({\bf r_c,R})$, for 
${\bf r_c}=(2,2)$, and (d) $S^{\omega}_2(R)$ (average of 
$S^{\omega}_2({\bf r_c,R})$ over ${\bf r_c}$).  Log-log ESS plots 
of the isotropic parts of the order-$p$ (e) velocity structure 
functions $S_p(R)$ versus $S_2(R)$ and (f) the vorticity 
structure functions $S_p^\omega(R)$ versus $S_2^\omega(R)$; 
$p=3$ (blue line with circles), $p=4$ (green line with triangles), 
$p=5$ (red line with squares), and $p=6$ (cyan line with stars); 
plots of the local slopes $\chi_p$ and $\chi_p^\omega$ (see text), in the 
forward-cascade inertial range: (e1) $\chi_p$ versus $\log_{10}S_2(R)$ and 
(f1) $\chi_p^\omega$ versus $\log_{10}S_2^\omega(R)$. 
Plots versus $p$ of the exponent ratios (e2) $\zeta_p/\zeta_2$, 
along with the KLB prediction (red line), and 
(f2) $\zeta_p^\omega/\zeta_2^\omega$ and error bars from the 
local slopes (see text). All plots are for run {\tt R7}.
} 
\label{fig:pomgstf}
\end{figure*}

Since Kolmogorov forcing is inhomogeneous, we use the 
decomposition $\psi = \langle \psi\rangle + \psi'$ and $\omega =
\langle \omega \rangle + \omega'$, where the angular brackets 
denote a time average and the prime the fluctuating part 
\footnote{Experiments~\cite{riv00,riv01,dan02} achieve 
homogeneity via a periodic, square-wave forcing with amplitude 
$F_0$; this introduces another time-scale in the problem; to 
avoid this complication we work with a time-indepedent force.}, 
to calculate the order-$p$ velocity and vorticity structure 
functions $S_p({\bf r_c,R})\equiv\langle|({\bf u'(r_c+R)-u'(r)})\cdot{\bf R}/R|^p \rangle$ 
and $S_p^\omega({\bf r_c,R})\equiv\langle|{\bf\omega'(r_c+R)-\omega'(r)}|^p\rangle$, 
respectively, where ${\bf R}$ has   magnitude $R$ and ${\bf r_c}$ is an origin. 
Figures~\ref{fig:pomgstf}(a) and (b) show pseudocolor plots of 
$S_2({\bf r_c,R})$ and $S_2^\omega({\bf r_c,R})$, respectively, for 
${\bf r_c}=(2,2)$; other values of ${\bf r_c}$ yield similar results so 
long as they do not lie near the boundary layer~(Table \ref{table:para}) of thickness 
$\delta_b$ (${\bf r_c}$ is chosen at least $5\delta_b$ away from all boundaries). 
We now calculate $S_2({\bf R})\equiv \langle S_2({\bf r_c,R}) \rangle_{{\bf r_c}}$ and 
$S_2^\omega({\bf R})\equiv \langle S_2^\omega ({\bf r_c,R}) \rangle_{{\bf r_c}}$, where 
the subscript ${\bf r_c}$ denotes an average over the origin (we use ${\bf r_c} = (i,j), 2 \leq i, j \leq 5$); these
averaged structure functions [Figs.~\ref{fig:pomgstf}(c) and (d)] are nearly isotropic 
for $R<\ell_{inj}$ but not so for $R >\ell_{inj}$.  To obtain the isotropic parts in an $SO(2)$ 
decomposition of these structure functions~\cite{bou05} we integrate over the angle $\theta$ that 
${\bf R}$ makes with the $x$ axis to obtain $S_p(R)\equiv\int_{0}^{2\pi} S_p({\bf R}) d\theta$ and
$S_p^\omega(R)\equiv \int_{0}^{2\pi} S_p^\omega({\bf R}) d\theta$.  
Given $S_p(R)$ and $S_p^\omega(R)$ we use the extended-self-similarity (ESS) procedure~\cite{ben93} to extract
the multiscaling-exponent ratios $\zeta_p/\zeta_2$ and 
$\zeta_p^\omega/\zeta_2^\omega$, respectively, from 
the slopes (in the forward-cascade inertial range) of log-log plots of 
$S_p(r)$ versus $S_2(r)$~[Fig.~\ref{fig:pomgstf}(e)] and 
$S_p^\omega(R)$ versus $S_2^\omega(R)$~[Fig.~\ref{fig:pomgstf}(f)] \footnote{We employ ESS since 
forward-cascade inertial ranges have a very modest extent even in the largest DNS studies~\cite{tsa05,bof07} 
that use periodic domains and hyperviscosity.}.  The insets Figs.~\ref{fig:pomgstf} (e1) and (f1) show, 
respectively, plots of the local slopes $\chi_p\equiv d\log_{10}S_p(R)/d\log_{10} S_2(R)$ versus $\log_{10} S_2(R)$ 
and $\chi_p^\omega\equiv d\log_{10}S_p^\omega(R)/d\log_{10}S_2^\omega(R)$ versus $\log_{10} S_2^\omega(R)$ 
in the forward-cascade regime; the mean values of $\chi_p$ and $\chi_p^\omega$, over the ranges shown,
yield the exponent ratios $\zeta_p/\zeta_2$ and $\zeta_p^\omega/\zeta_2^\omega$ that are plotted versus $p$ in 
Figs.~\ref{fig:pomgstf} (e2) and (f2), respectively, in which the error bars indicate the maximum deviations 
of $\chi_p$ and $\chi_p^\omega$ from their mean values. The Kraichnan-Leith-Batchelor (KLB) predictions~\cite{kraic67} 
for these exponent ratios, namely, $\zeta^{KLB}_p/\zeta^{KLB}_2 \sim r^{p/2}$ and 
$\zeta_p^{\omega,KLB}/\zeta_2^{\omega,KLB} \sim r^0$, agree with our values for $\zeta_p/\zeta_2$ but not
$\zeta_p^\omega/\zeta_2^\omega$: velocity structure functions do not display multiscaling~[Fig.~\ref{fig:pomgstf} (e2)] 
whereas their vorticity analogs do~[note the curvature of the plot in Fig.~\ref{fig:pomgstf} (f2)]. 
This is in consonance with the results of DNS 
studies with periodic boundary conditions~\cite{tsa05,bof07}. Indeed, if we use the same values of 
$\gamma$ as in Ref.~\cite{tsa05}, we obtain the same exponent ratios (within error bars); thus our method for the
extraction of the isotropic parts of the structure functions suppresses boundary and anisotropy effects efficiently.

\begin{figure}[!htb]
\includegraphics[width=0.88\linewidth]{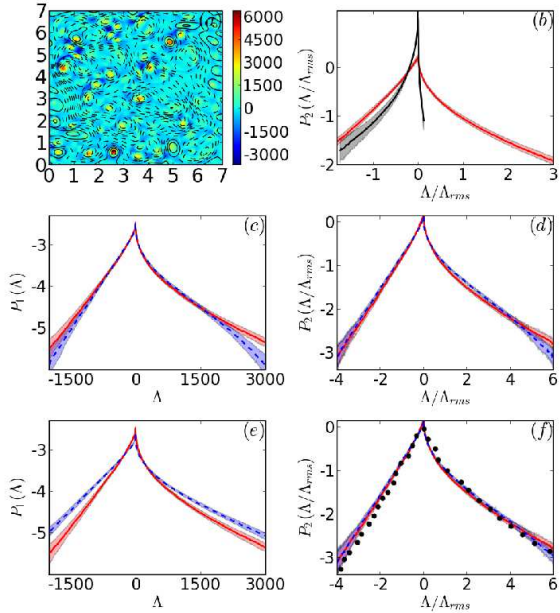}
\caption{\small (Color online) (a) Representative pseudocolor 
plot of $\Lambda$  superimposed on a contour plot of the stream 
function $\psi$; (b) the PDF $P_1(\Lambda)$ obtained from points in 
the bulk $\delta_b < x,y < L-\delta_b$ (red line) and from 
those points within a distance $\delta_b$ from  the boundaries 
(black line) for our run ${\tt R7}$; plots of 
(c) $P_1(\Lambda)$ versus $\Lambda$ and (d) $P_2(\Lambda/\Lambda_{rms})$ 
versus $\Lambda/\Lambda_{rms}$ for fixed ${\mathcal G}$ and 
$\gamma=0.25$(red line) and $\gamma=0.71$(blue dashed line)  
[runs ${\tt R4}$ and ${\tt R5}$]; plots of (e) $P_1(\Lambda)$ 
versus $\Lambda$ and (f) $P_2(\Lambda/\Lambda_{rms})$ versus 
$\Lambda/\Lambda_{rms}$  [runs ${\tt R4}$ and ${\tt R6}$ with  
$Re\simeq23.5$] and $\gamma=0.25$(red line) and 
$\gamma=0.71$(blue dashed line) and points (black dots) extracted from 
Fig.~2(d) of Ref.~\cite{riv01}. In (c)-(f) the fluctuating part of the 
velocity is used for $\Lambda$. One-standard-deviation error bars 
are indicated by the shaded regions.} 
\label{fig:pdflam}  
\end{figure}

For an inviscid, incompressible 2D fluid the local flow topology 
can be characterized via the Weiss criterion~\cite{wei92} that 
uses the invariant $\Lambda\equiv(\omega^2 - \sigma^2)/8$, where 
$\sigma^2\equiv\sum_{i,j} \sigma_{ij}\sigma_{ji}$ and 
$\sigma_{ij}\equiv (\partial_i u_j + \partial_j u_i)$. This 
criterion provides a useful measure of flow properties even if 
$\nu > 0$ as noted in the experiments of Ref.~\cite{riv01}: 
Regions with $\Lambda > 0$ and $\Lambda < 0$ correspond to 
centers and saddles as we show in Fig.~\ref{fig:pdflam} (a) by 
superimposing, at a representative time, a pseudocolor plot of 
$\Lambda$ on contours of $\psi$. This result is in qualitative 
accord with experiments [see, e.g., Fig.~$1$ of 
Ref.~\cite{riv01} and also earlier DNS studies~\cite{wei92}, 
which do not use Ekman friction].  In Fig.~\ref{fig:pdflam} (b) we 
compare the scaled PDFs $P_2(\Lambda/\Lambda_{rms})$ with data 
obtained from points near the walls (black curve) and from points 
in the bulk (red curve); the clear difference between these, not 
highlighted before, indicates that regions of large  
$\Lambda$ are suppressed in the boundary layers. There is a
generation of strain and vorticity in these boundary layers and 
scatter plots, not shown, indicate $\omega^2 \simeq \sigma^2$ 
here; this leads to the suppression of regions of large  
$\Lambda$ in $P_2(\Lambda/\Lambda_{rms})$.

\begin{figure}[!htb]
\includegraphics[width=0.95\linewidth]{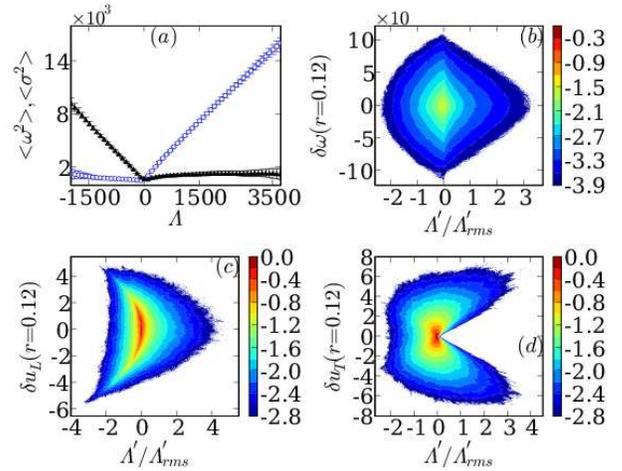}
\caption{
(Color online)
(a) Plots of conditional expectation values, with one-standard-deviation 
error bars, of  $\langle\sigma^2\rangle$(dots) and $\langle\omega^2\rangle$(circles) 
for a given $\Lambda$; pseudocolor plots of (b) the joint PDF 
$P(\delta \omega(r=0.12),\Lambda^\prime/\Lambda_{rms}^\prime)$, 
(c) the joint PDF $P(\delta u_L(r=0.12),\Lambda^\prime/\Lambda_{rms}^\prime)$, and 
(d) the joint PDF $P(\delta u_T(r=0.12),\Lambda^\prime/\Lambda_{rms}^\prime)$ for 
our run {\tt R7}. The contours and the shading are for the logarithms 
of the joint PDFs.}
\label{fig:conlam}
\end{figure}

Figures~\ref{fig:pdflam} (c) and (d) show the PDF $P_1(\Lambda)$ 
and the scaled PDF $P_2(\Lambda/\Lambda_{rms})$ for runs ${\tt R4}$ (red line) 
and ${\tt R5}$ (blue dashed line), with $\gamma=0.25$ and $\gamma=0.71$, 
respectively, and ${\mathcal G}=3.5\times 10^4$; 
by comparing these figures we see that both $P_1$ and $P_2$ overlap 
within error bars for runs ${\tt R4}$ and ${\tt R5}$.  We believe this 
is because, in fixed-${\mathcal G}$ runs like ${\tt R4}$ and ${\tt R5}$, 
$\epsilon_{\nu}$ does not change [Table~\ref{table:para}] even though $\gamma$ changes.  
By contrast, if we compare $P_1$ and $P_2$ [Figs.~\ref{fig:pdflam} (e) and (f)] 
for runs ${\tt R4}$ (red line) and ${\tt R6}$ (blue dashed line), in which 
the mean $Re$ is held fixed by tuning both $\gamma$ and ${\mathcal G}$, we find, in agreement with
experiments~\cite{riv01}, that the PDFs $P_1$ do not agree for 
these runs, but the PDFs $P_2$ overlap within error bars. Our 
results for $P_2$ in Fig.~\ref{fig:pdflam}(f) are in quantitative 
agreement with experiments: we have obtained the points in this 
plot by digitising data points [see http://www.frantz.fi/software/g3data.php]  
in Fig.~2(d) of Ref.~\cite{riv01}; the errors in 
these points are comparable to the spread of data 
in~\cite{riv01}.  Conditional expectation values of 
$\langle\sigma^2\rangle$ and $\langle\omega^2\rangle$, for a 
given value of $\Lambda$, also agree well with experiments as can 
be seen by comparing Fig.~\ref{fig:conlam} with Fig.~$3$ of 
Ref.~\cite{riv01}.  We also present in Figs.~\ref{fig:conlam} (b-d) 
pseudocolor plots of the joint PDFs of 
$\delta \omega(r) \equiv \omega^\prime({\bf x} + r \hat{e}_x) - 
\omega^\prime({\bf x})$,  $\delta u_L(r) \equiv u_x^\prime({\bf x} + r \hat{e}_x) - u_x^\prime({\bf x})$ or $\delta u_T^\prime(r) 
\equiv u_y^\prime({\bf x} + r \hat{e}_x) - u_y^\prime({\bf x})$ with 
$\Lambda^\prime \equiv \det(M)$, $M^{\alpha \beta} \equiv 
\int_\Omega m^{\alpha \beta} \, {\mathbf dr}/{\int_\Omega 
{\mathbf dr}}$, $m^{\alpha \beta} \equiv \partial_\alpha {\bf 
u}_\beta^\prime$, $\Omega$ a circular disc with center at 
${\bf x} + (r/2)\hat{e}_x$ 
and radius $r/2$, and $r$ in the forward-cascade regime; we obtain striking 
agreement with experiments as can be seen by comparing 
Figs.~\ref{fig:conlam} (c-d) with Figs.~$1$-$2$ of Ref.~\cite{dan02}.  
Finally, we calculate $\tilde{\Lambda}\equiv\sqrt{\langle(\partial_x u_y^\prime)^2\rangle
\langle(\partial_y u_x^\prime)^2\rangle}$ and $b\equiv-\langle
\partial_xu_y^\prime \partial_y u_x^\prime \rangle/\tilde{\Lambda}$ 
(see Table \ref{table:para}) and obtain excellent 
agreement with experiments \cite{riv01}. 

Some earlier numerical studies of 2D, wall-bounded, statistically 
steady turbulent flows~\cite{cle05} use forcing functions that are not of 
the Kolmogorov type; furthermore, they do not include Ekman 
friction.  Other numerical studies, which include Ekman friction 
and Kolmogorov forcing, employ periodic boundary 
conditions~\cite{ban08,bof07,tsa05}.  To the best of our 
knowledge our study of 2D {\it turbulent} flows is the first one 
that accounts for Ekman friction, realistic boundary conditions, 
and Kolmogorov forcing. Thus we can make quantitative comparisons 
with soap-film experiments; and the agreement between our results 
and those of Refs.~\cite{riv00,riv01,dan02,riv07} vindicates the 
use Eq.~(\ref{eq:ns_vor}) as a model for these soap films~\cite{cho01}. 
We hope our results will stimulate experimental studies designed 
to extract (a) the isotropic parts of structure functions (and thereby to 
probe the multiscaling of vorticity structure 
functions~[Fig.~\ref{fig:pomgstf} (f1)]) or (b) the 
PDF $P_2(\Lambda/\Lambda_{rms})$~[Fig.~\ref{fig:pdflam} (b)] near soap-film boundaries.

We thank J.~Bec, G. Falkovich, V.~Kumar, D.~Mitra, and S.S.~Ray for 
discussions, CSIR, DST, and UGC(India) for financial support, and 
SERC(IISc) for computational facilities. RP is a member of the 
International Collaboration for Turbulence Research.

\end{document}